\documentclass[sigconf]{acmart}
\pdfoutput=1
\usepackage{mathptmx} 
\usepackage{newtxmath}
\usepackage{siunitx}
\usepackage{fancyhdr}
\usepackage{bbding}
\usepackage{makecell} 
\usepackage{graphicx}
\usepackage{subfig}
\usepackage{url}
\usepackage{algorithm}
\usepackage{algorithmic}
\usepackage[normalem]{ulem}
\renewcommand\footnotetextcopyrightpermission[1]{}

\begin{document}

\title{GTA: a new General Tensor Accelerator with Better Area Efficiency and Data Reuse}

\author{Chenyang Ai}
\affiliation{
  \institution{Peking University}
  \city{Beijing}
  \country{China}}
\email{chenyang_ai@stu.pku.edu.cn}

\author{Lechuan Zhao}
\affiliation{
  \institution{Peking University}
  \city{Beijing}
  \country{China}}
\email{lczhao@stu.pku.edu.cn}

\author{Zhijie Huang}
\affiliation{
  \institution{Peking University}
  \city{Beijing}
  \country{China}}
\email{zhijiehuang@stu.pku.edu.cn}

\author{Cangyuan Li}
\affiliation{
  \institution{Chinese Academy of Sciences}
  \city{Beijing}
  \country{China}}
\email{licangyuan20g@ict.ac.cn}

\author{Xinan Wang}
\affiliation{
  \institution{Peking University}
  \city{Beijing}
  \country{China}}
\email{anxinwang@pku.edu.cn}

\author{Ying Wang}
\affiliation{
  \institution{Chinese Academy of Sciences}
  \city{Beijing}
  \country{China}}
\email{wangying2009@ict.ac.cn}

\begin{abstract}
Recently, tensor algebra have witnessed significant applications across various domains. Each operator in tensor algebra features different computational workload and precision. However, current general accelerators, such as VPU, GPGPU, and CGRA, support tensor operators with low energy and area efficiency. This paper conducts an in-depth  exploration of general accelerator for tensor processing.

First, we find the similarity between matrix multiplication and precision multiplication, and create a method classifying tensor operators using GEMM and vector operations. Then, we implement two discoveries and introduce the systolic architecture into general-purpose accelerator. Therefore, we propose a new General Tensor Accelerator (GTA), which has a better area efficiency and data reuse. Furthermore, we create a general tensor scheduling optimization strategies based on dataflow, precision and array resize. Our evaluation results demonstrate that GTA is able to achieves $7.76\times$, $5.35\times$, $8.76\times$ memory efficiency and $6.45\times$, $3.39\times$, $25.83\times$ speedup over of VPU, GPGPU and CGRA.
\end{abstract}


\maketitle

\section{Introduction}
In recent years, computing kernels in various domains, such as machine learning, data analysis, signal processing, and scientific computing, have been categorized as operators in tensor algebra (like GEMM, CONV, GEMV, MTTKRP, TTMc etc.). These tensor operators exhibit different computational workload, including inherent variation in tensor dimension, computation, and accumulation. ~\cite{jia2022automatic,lu2021morphling}. Furthermore, these operators involve different precision. For instance, applications like audio and image signal processing often demand filter operators with INT8 or INT16 precision~\cite{spanias2006audio,widrow2008quantization}. Scientific computing, encryption algorithms, and other zero-error algorithms necessitate NTT(Number Theoretic Transform) with INT32 or even INT64 precision~\cite{knuth2014art,bailey2012high,zhai2022accelerating}. Even in one domain, there can be diverse precision requirements. In machine learning, for example, quantization inference (INT8)~\cite{jacob2018quantization}, training weights (FP16, BP16)~\cite{henry2019leveraging}, and high-precision AI networks (FP32, FP64)~\cite{bi2023accurate} encompass a wide spectrum of precision. 

Previous works focus on dedicated ASIC accelerators for specific tensor operator~\cite{chen2016eyeriss,choquette2023nvidia,jia2022automatic}. But, the general purpose hardware still have many application scenarios, and they are typically employed to handle various tensor applications. With overwhelming computation, CPU can't bear the burden of tensor operators. Existing general accelerators adopted by academia and industry refer to Vector Processing Unit (VPU) and Cuda Core in GPGPU. 

However, these vector architectures support the computational workload of operators with low energy efficiency~\cite{casper2014hardware,pedram2012codesign}. One of the most important reasons is that vectorization is indeed the most general method for tensor operators, but the computing unit cannot exploit data reuse in tensor operators, resulting in a lot of access to storage. 

Besides, the emergence of various precision of tensor operator requires hardware support. For example, contemporary vector instruction sets need to implement operations for eight kinds of precision, including INT8, INT16, INT32, INT64, BP16, FP16, FP32, and FP64~\cite{armsve,rvvspec,IntelAVX-512}. These precision units are often established independently and occupy the majority of the overall area~\cite{Ara2020}. But, the specific workload always only utilizes single precision unit of all, which leads to a low area efficiency.

To summarize, current general accelerators suffer from low area and energy efficiency simultaneously. Actually, it is difficult to achieve both generality and efficiency in hardware design. To our knowledge, previous works only improves the energy efficiency of GEMM~\cite{ju202265nm,VSA,SMA} for general accelerators. There are also solutions to introduce CGRA~\cite{lu2021morphling} into general tensor processing. However, their energy efficiency and generality can be further improved, and they overlook opportunity to co-optimize with computational precision.

This work takes an ingenious perspective, dissecting the computational workload and precision from the perspective of architecture and computing paradigms. Based on traditional VPU, we proposes a new General Tensor Accelerator (GTA) aiming for processing tensor applications more efficiently. In summary, this paper makes the following contributions:\par
\begin{itemize}
 \item We find the similarity between matrix multiplication and precision multiplication, and create a classification of tensor operators. Then we implement two discoveries on systolic array for better area and memory efficiency.
 \item We design a Multi-Precision Reconfigurable Array (MPRA) and implement MPRA in vector architecture to compose GTA, which can compute the tensor operators with arbitrary computational workload and precision. 
 \item We implement general tensor scheduling optimization strategies based on dataflow, precision and array resize and make an analysis of scheduling space.
 \end{itemize}
 According to the evaluation, GTA is able to achieve $7.76\times$, $5.35\times$, $8.76\times$ memory efficiency and $6.45\times$, $3.39\times$, $25.83\times$ speedup over VPU (Ara),  GPGPU (NVIDIA H100) and CGRA (hycube).

\section{BACKGROUND}
In this section, We provide here an introduction to the specialized accelerators about tensor operators compared with general accelerators. The accelerators are mainly spatial architecture, and systolic array is a typical example. Then we provide a brief overview of previous work about reconfigurable architecture and reveal the distance to a general tensor accelerator.

\begin{figure}[htbp]
\centerline{\includegraphics[width=1.1\linewidth]{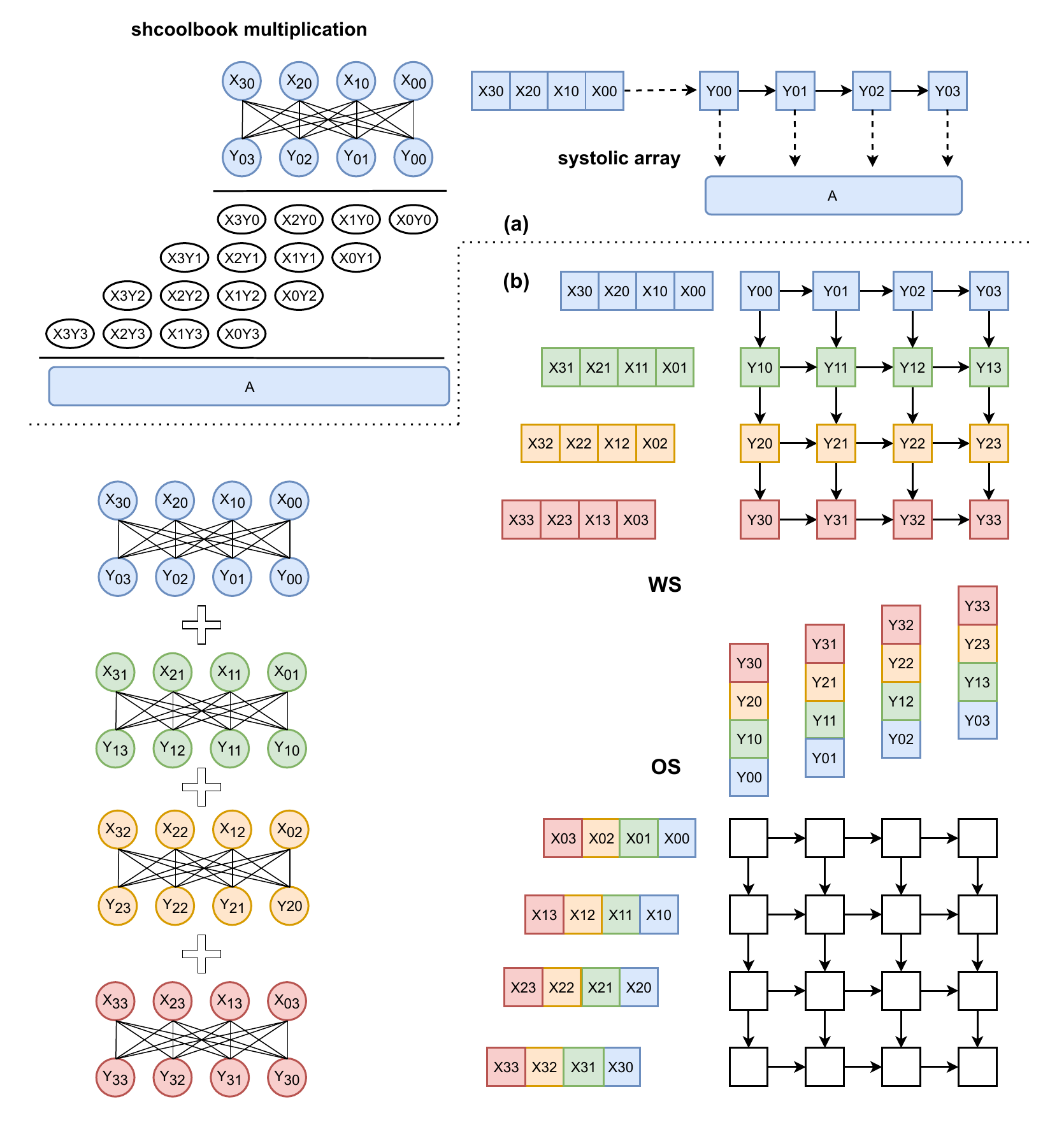}}
\caption{The diagram of multi-precision matrix multiplication implement on the systolic array.}
\label{schoolbook multiplication}
\end{figure}

\subsection{Specialized Accelerators and Systolic Array}
\label{Specialized}
To tackle the problem of tensor algebra acceleration, various spatial architectures are adopted such as Tensor Core~\cite{choquette2023nvidia}, eyeriss~\cite{chen2016eyeriss}, etc. These spatial accelerators are often composed of PE arrays and their interconnections, which can exploit different types of data reuse. Usually, the majority of spatial architectures is naturally suitable for single kind of tensor operator with specific computational workload because of their microarchitecture. For example, the systolic array is well suited to accelerating GEMM, while eyeriss is for CONV. In most cases, they are only designed for one precision.

Among them, the systolic array has emerged as a great choice due to its high power efficiency and peak throughput. Adopting simple PE unlike that with large memory in other dataflow architectures~\cite{chen2016eyeriss} and implementing simple interconnection instead of complicated and poor timing one in Tensor Core~\cite{choquette2023nvidia} makes it easy to scale the array, which achieves a higher degree of data reuse~\cite{yuzuguler2023scale,samajdar2020systematic}. Furthermore, there are several works on improving the utilization, which is the primary drawback of the systolic array. RASA~\cite{jeong2021rasa} utilizes systolic array as an additional pipeline matrix unit of CPU, also overlooking the potential for reusing the original components. Mirroring~\cite{Dataflow_mirroring} and Redas~\cite{ReDas} work on flexible multi-workloads to improve utilization. All the above works have not jointly optimized dataflow and precision. Meanwhile, previous efforts have integrated systolic array with CPU~\cite{ju202265nm}, VPU~\cite{VSA}, GPU~\cite{SMA}. But they cannot fully exploit original control and communication hardware resources, which could have been used to improve the utilization of systolic array. They also ignore the precision unit of general accelerators.

\subsection{Specialized Accelerators Become More and More General}
Recently, while general-purpose accelerators are enhancing their performance for GEMM in neural networks~\cite{ju202265nm,VSA,SMA}, specialized accelerators are trying to accept more operators in robustness. But we hold the view that previous works only explore reconfigurable architecture and functional versatility, which involves limited range of computational workload and precision. 

For example, there are a few studies exploring the functional versatility of TPU~\cite{li2023append,gptpu}. These works lack hardware modifications, resulting in limited improvement. Furthermore, they don't take an intrinsic view into the generality of tensor operators. Morphling~\cite{lu2021morphling} addresses how to transform CGRA to accommodate dense and sparse tensor operators. We think that CGRAs often require large-area interconnect, memory and control units. Therefore the hardware often comprises small array, like $8\times 8$ and $4\times 4$, with inadequate ability of data reuse and accelerating. Besides, critical precision aspect is overlooked. Tensorlib~\cite{jia2022automatic} focuses on the generation of hardware towards tensor operators. But it ignores the generality and leads to under-utilization of the array.

Above works all ignore the chance to co-optimize with the precision, which is the significant factor about implementation. Bitfusion~\cite{sharma2018bit} and GPNPU~\cite{song2020gpnpu} consider the precision and computation at the same time, but are stuck in the limited range of computational precision and workload domains of operators. 

\section{INSIGHTS INTO MULTI-PRECISION TENSOR OPERATORS}

In this section, we analyze the inherent characteristics of the computational workload and precision of tensor operators. We find the similarity between Matrix Multiplication and precision multiplication and a method classifying tensor operators. At last, we implement two discoveries on systolic array in each subsection respectively.

\subsection{Similarity between Matrix Multiplication and Precision Multiplication}
\label{similarity}

 It is known that a long multiplication could be decomposed
into several basic multiplications and additions. As shown in Figure \ref{schoolbook multiplication}(a), operands (i.e.X and Y are 32 bits) are decomposed into 8-bit limbs respectively (like $ X/_{00} $,$ Y/_{00} $, etc). It is obvious that all 8-bit limbs of the X and Y need to be cross-multiplied, and then the product of limbs should be added to get the result, which is similar to the computing pattern in matrix multiplication (by the way, the carry-bits among the product of limbs will be processed in the accumulator).

The reuse pattern of systolic array is in conformity with matrix multiplication, which can be migrated to multiplication detailed above. It could assume that the precision of a single PE is 8 bits. As can be seen from Figure \ref{schoolbook multiplication}(a), in Weight-Stationary (WS) mode, we can place the limbs in consecutive positions as the weight used in WS mode and the other one is taken as input streaming into the array from the leftmost PE. And 32-bit multiplication is achieved within 4 PEs.

As shown in Figure \ref{schoolbook multiplication}(b) of WS mode, in the column direction, the partial product of this multiplication flows downward to next row. Therefore the corresponding partial products produced at the same position are added, which is essentially equivalent to the addition of two multiplication results. Also, the input could be reused by row direction. After extending the row and column, it is equivalent to a multi-precision matrix multiplication. 

The three types of dataflow supported by the systolic array in this work are WS, Input-Stationary(IS), and Output-Stationary(OS). The dataflow of IS is the same as that of WS, and the operands occupying the array are inputs. As shown in Figure \ref{schoolbook multiplication}(b) of OS mode, the difference between multi-precision in OS and WS is analogous to the distinctions in matrix multiplication. However, in OS mode, due to the fact that both input and weight are mapped onto the array, the size of the workload mapped on the array expands with multiple in both the column and row directions. In contrast, when working in WS mode, it only affects the row direction. Leveraging the array's scalability, it could enable the realization of matrix multiplication with arbitrary multiples of PE's precision.

\subsection{Classification of Tensor Operators using p-GEMM and vector operators}

\begin{figure}[htbp]
\centerline{\includegraphics[width=0.7\linewidth]{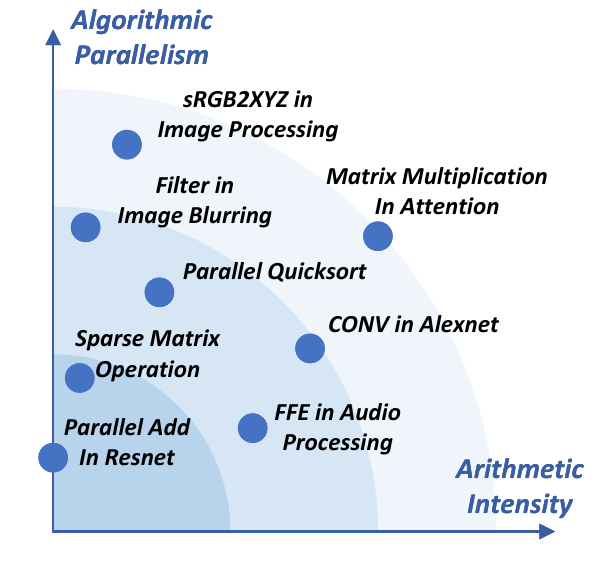}}
\caption{The indicative algorithmic parallelism and arithmetic intensity of some tensor operators.}
\label{tensor operator}
\end{figure}

It is worth reconsidering these tensor operators from a hardware perspective. The computational workload of tensor operators can be decomposed into two dimensions: algorithmic parallelism and arithmetic intensity. Arithmetic intensity reflects the data reuse opportunity of operators, indicating how many computations are performed after fetching data from memory. Algorithmic parallelism represents the parallel operations that could be extracted under certain circumstances, like the extent of vectorization achievable in a kernel. For instance, not absolutely, image processing algorithms always exhibit higher algorithmic parallelism compared to audio processing algorithms. These two dimensions serve as axes for partitioning existing tensor operators, yielding the following approximate classification results in Figure \ref{tensor operator}.

Along the arithmetic intensity axis, tensor operators with no intensity could only be compiled into vector operations without data reuse opportunity, while those with higher intensity could be transformed into GEMM, like solving tensor contraction problems. Tensor contractions can be rewritten equivalently as the form of Transpose-Transpose-GEMM-Transpose sequences~\cite{gett,progressively_raising}. Also, it is easy to manually convert to a GEMM kernel based on the data reuse characteristics.
Based on the results of the above works, these GEMM come with different sizes, whose transformation results depend on both the arithmetic intensity and algorithmic parallelism, encompassing operations of varying size hierarchies such as matrix multiplication, matrix-vector multiplication, and vector inner product. In general, operators with high algorithmic parallelism and arithmetic intensity are transformed into matrix multiplication.

Therefore, we can define them as p-GEMM (p represents pseudo) including operators of arbitatry size. Although systolic dataflow architecture is ideal for handling matrix multiplication, it may lead to under-utilization problems when meeting various operators in p-GEMM unsuitable for the size of array, let alone the vector operations. Vector hardware has flexible interconnection and control units. So we could combine the vector and systolic architecture to improve the utilization of systolic array by reusing these fine-grained control modules. We will discuss this in detail in the next section.\par

\section{HARDWARE ARCHITECTURE}

\begin{figure}[htbp]
\centerline{\includegraphics[width=0.6\linewidth]{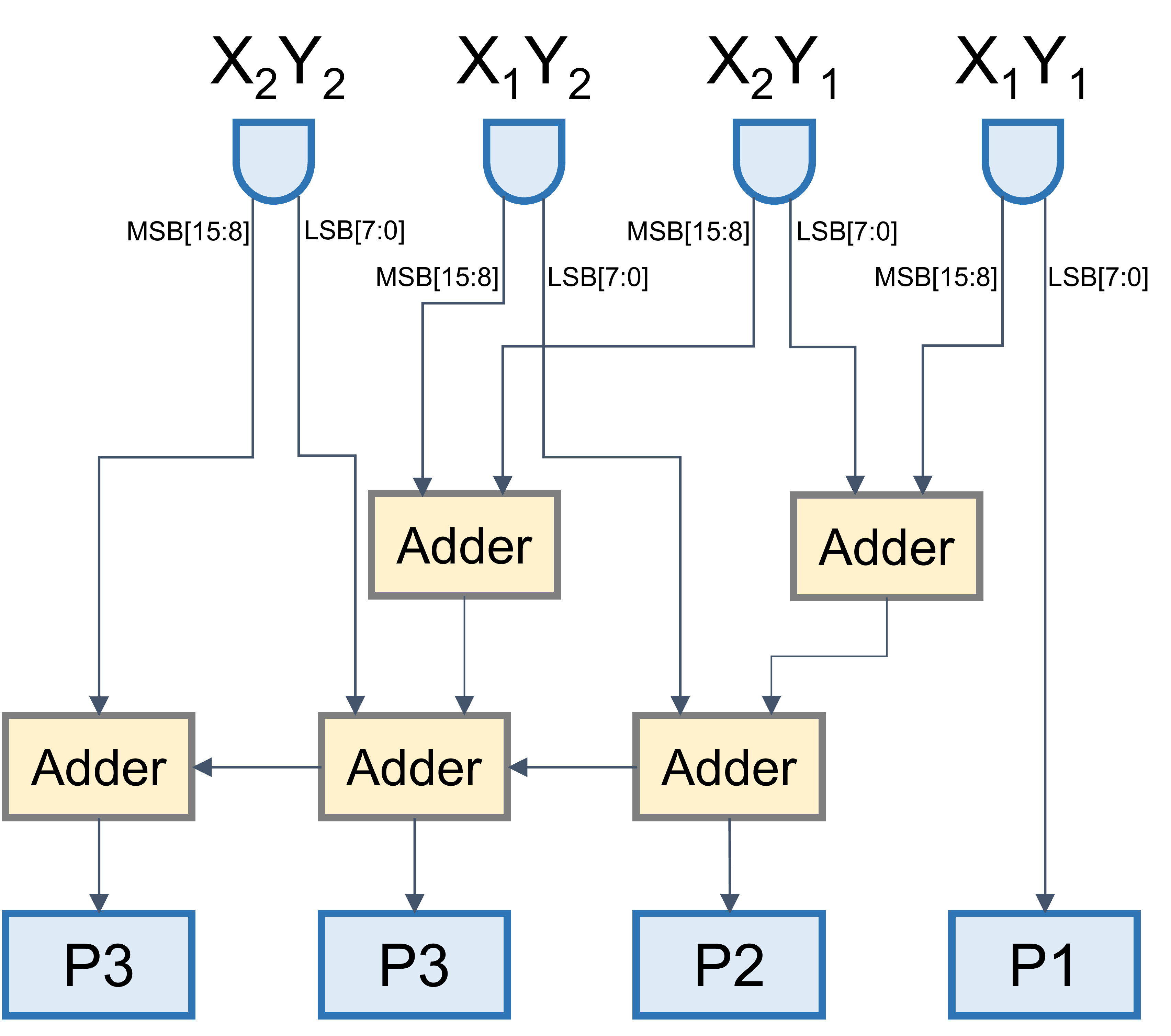}}
\caption{Implementation of a 16-bit multi-precision accumulator.}
\label{adder}
\end{figure}

\begin{figure*}[htbp]
\centerline{\includegraphics[width=7.2in]{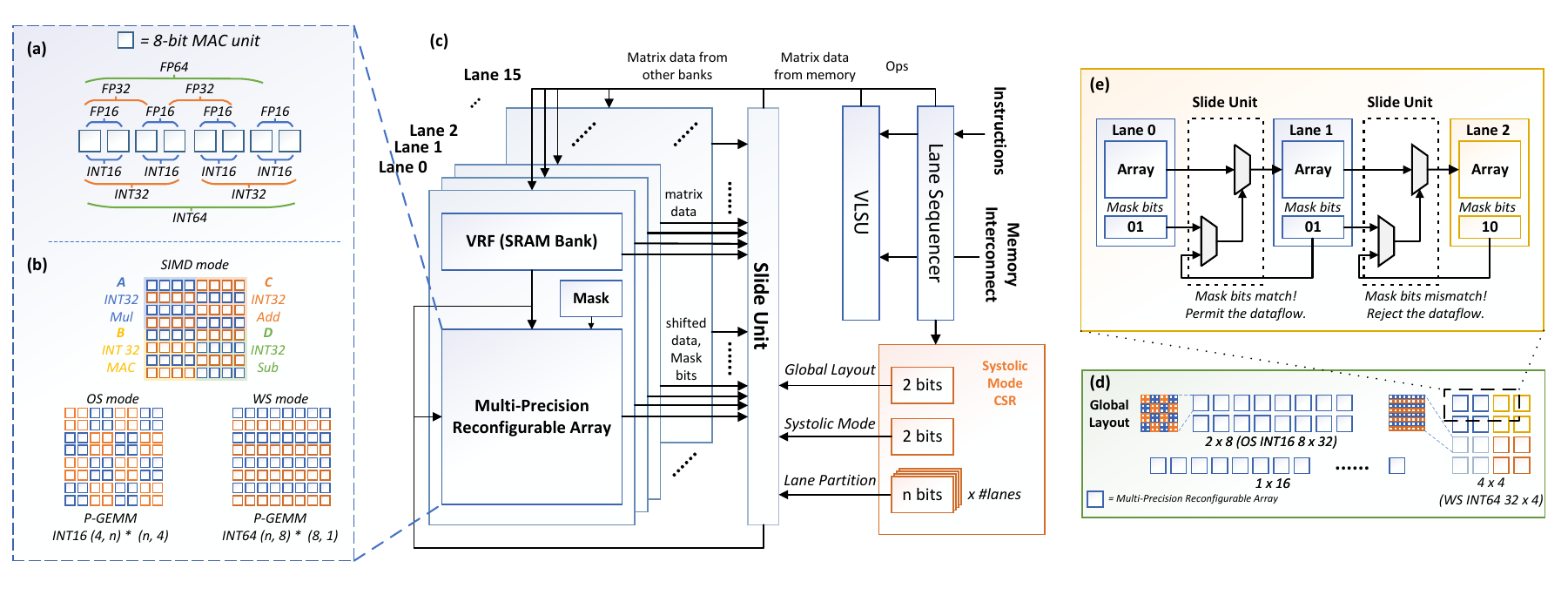}}
\caption{Architecture overview: example of 16 lanes. (a) One row of MPRA cover various kinds of precision in WS mode. (b) $8\times 8$ MPRA operate multi-precision p-GEMM and vector operation. (c) The overview of GTA architecture (d) The MPRA combine the whole array with reconfigurable shape (e) The modified slide unit use mask bits to arrange the dataflow of array.}
\label{architecture}
\end{figure*}
We choose the VPU as the combining vector hardware. In this section, based on above discoveries, the GTA architecture will be explained, detailing the implementation of Multi-Precision Reconfigurable Array (MPRA) and reuse of fine-grained control and interconnection logic in VPU. 

\subsection{Multi-Precision Reconfigurable Array}\label{Similarity}

We can call it Multi-Precision Reconfigurable Array(MPRA) which could implement systolic dataflow to operate multi-precision matrix multiplication as described in Section \ref{similarity}. As shown in Figure~\ref{architecture}(a), the precision that a single PE supports is 8-bit and the multiplication is performed in a linear pattern. The number of column is set to 8 so that one row of PEs supports the $8\times n$ bits ($n=1,2,4,8$) multiplication in WS or IS mode by utilizing corresponding partition. A MPRA set 64($8\times 8$) PEs to support 64-bit in OS mode. 

The multi-precision accumulator is composed of basic accumulator units to support accumulation in different bit width. We take a 16-bit accumulator as an example. Figure \ref{adder} shows the microarchitecture of a basic 16-bit accumulator unit. As shown, a 16-bit accumulator unit takes as input four 16-bit operands – $X_1Y_1$, $X_2Y_1$, $X_1Y_2$ and $X_2Y_2$, which corresponds to four partial products of 16-bit multiplications generated by systolic array. Based on the mathematical property, the 16-bits accumulator unit uses shift-add operations to easily generate the results of 16-bit multiplications.

MPRA can be reconfigured to perform mantissa multiplication in different width, coordinated with other functional units to execute the FP operation. In addition to mantissa computation, the FPadd and FPmul require alignment, normalization, overflow processing, rounding and other steps. And the dominant area and energy consumption comes with the multiplier of the mantissa. Specifically, the mantissa multiplication for BP16, FP16, FP32, and FP64 can be equivalently represented as the multiplication of INT8, 12, 24, and 53, respectively.

\subsection{GTA Overall Architecture}

As can be seen in Figure \ref{architecture}(c), the VPU could be configured with various lanes and a lane scheduler responsible for inter-lane communication and allocation of computational resources. Within each lane of original VPU, Multiply Accumulate (MAC) units in various precision are set up to meet the requirements of the vector extension instruction set, occupying a substantial area. We introduce one MPRA into each lane to replace these MAC units. Due to the configurability of GTA architecture, we can set arbitrary number of lanes. We take an example of 16 lanes below.

Besides the MAC unit, the PE in MPRA is equipped with three operand registers, systolic mode register, operation units (the same as lane's), and a centrally controlled finite state machine. The systolic mode register is synchronized with the global configuration in CSR, which controls the data transfer of single PE. \par

We introduce the Systolic Control and Status Register (SysCSR) as shown in Figure~\ref{architecture}(c), which achieves a three-level array interconnect configuration. These levels correspond to the Global Layout, Systolic Mode, and Mask Group fields, respectively. The Global Layout field encodes the logical layout of lanes, and upon decoding, programs direction of the data shuffling (source lane and destination lane) in the Slide Unit as shown in Figure~\ref{architecture}(d). We can define it as array arrangement. This influences the interconnection of the lane, thus inducing different shapes of the array. The Systolic Mode specifies the content when data movement is performed between lanes. For example, in the GEMM-OS mode, the movement with three sets of operands between lanes is required, while in the GEMM-WS(IS) mode, a set of input data and partial sum results need to be transferred. The situations in IS mode is similar to that of WS mode.\par

 \begin{figure*}[htbp]
\centerline{\includegraphics[width=6in]{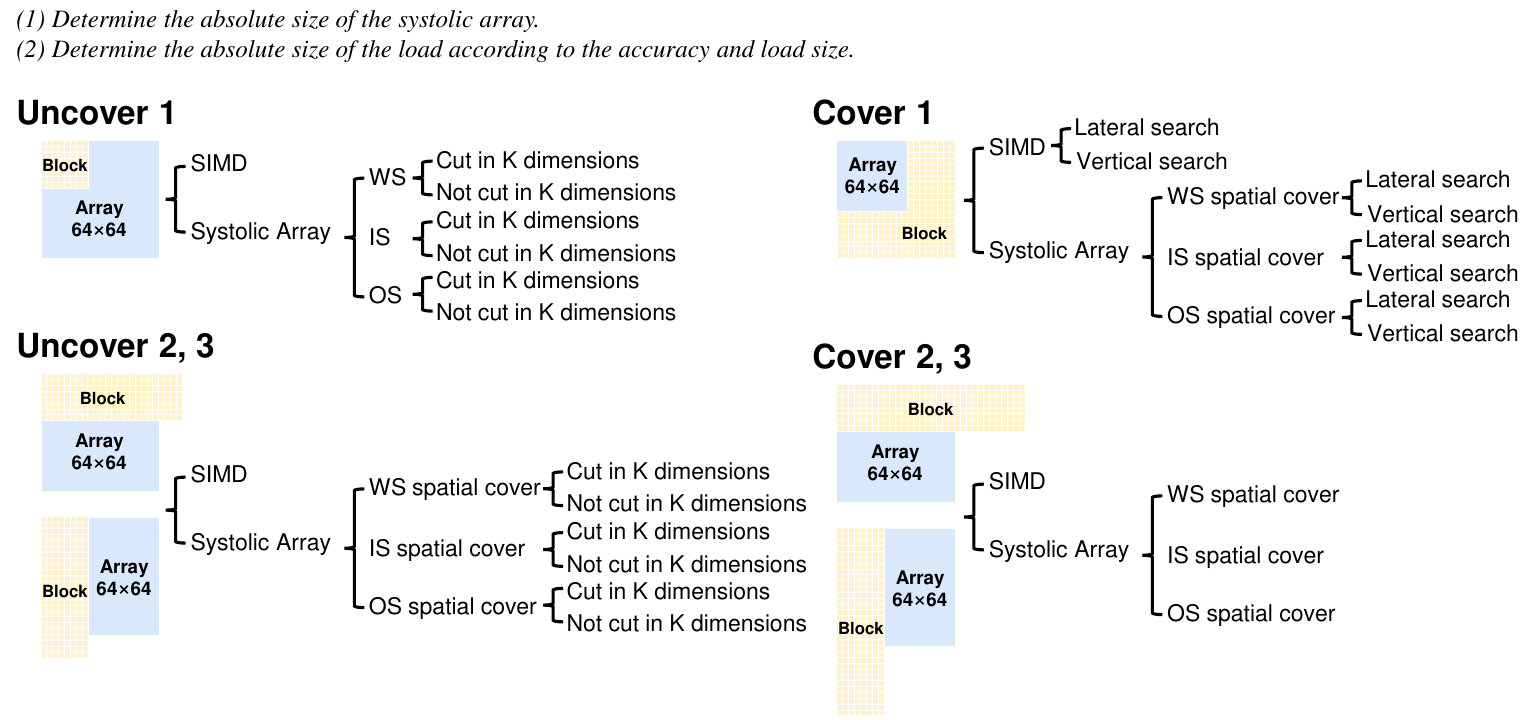}}
\caption{Dataflow pattern matching: 64 lanes and $64\times64$ size array example.}
\label{classify}
\end{figure*}

The Mask Sets consist of a group of mask bit sets in the number of lanes. After operator scheduling is completed, the hardware library generates mask bit sets based on shape information. Following the execution of SysCSR write instructions, the Lane Scheduler loads these sets into the mask registers within each lane to control the data transfer. As shown in Figure~\ref{architecture}(e), we introduce the Mask Match Mechanism to limit the data communication between lanes to logically divide lanes into different sub-regions, each of which contains lanes possessing a same set of mask bits permitting the data transfer. And the width of mask bits determines how many partitions are achievable in the architecture. Therefore GTA could combine its all MPRA as a whole array with several array rearrangements and freely schedule matrix operation of arbitrary size in high array utilization. \par

In conclusion, by reconfiguring the interconnection between PEs, MPRA can perform in either vector operator supported by original VPU or p-GEMM operator in multi-precision as shown in Figure~\ref{architecture}(a) and (b). In other words, the previous works~\cite{ju202265nm,VSA,SMA} only equip accelerator with GEMM and vector operation, while our work evolve into p-GEMM. The proposed architecture is applicable to a wide range of scenarios according to the number of configuring lanes, spanning from high-performance general computing to embedding accelerator. Compared with popular products in academia and industry, "VPU+Systolic Array" system~\cite{jouppi2021ten,fell2021marenostrum,xue2023v10}, GTA improve area efficiency in both aspects of computational precision and combining vector and matrix unit.\par

\section{SCHEDULING SPACE EXPLORING FOR p-GEMM OPERATOR}

The vector operators are executed by GTA as usual VPU. However, for a p-GEMM operator, the scheduling approach is influenced by three factors, including the array resize, computational precision, dataflow. According to section \ref{similarity}, the size mapped on the GTA depends on the computational workload of the tensor operator transformed into p-GEMM and its corresponding precision. Also, the mapping size is influenced by selected dataflow. We employ three systolic dataflows (WS, IS, OS) and SIMD(vector) mode (some p-GEMM operators may get better result from vectorization) to address this challenge. Additionally, the shape of whole array depends on array resize with numerous lanes. Different p-GEMM operators benefit from different array shape.

 In the context of matrix multiplication in three systolic dataflows, typically characterized by three dimensions: $M$, $N$, and $K$, where $M$ and $N$ can be assumed as two dimensions mapped onto the array spatially, and $K$ represents the temporal dimension of the operation.

 To enhance array utilization, the left computational tasks are mapped to the available part of array. In such cases, we can segment the $K$ dimension to increase the array utilization to reduce computation time. Moreover, according to the scale-sim ~\cite{samajdar2020systematic}, larger continuous computation areas mapped on the systolic array result in a higher degree of data reuse. Conversely, if the segmentation is thinner, memory access counts increase. Therefore, in these situations, the theoretical conflict between improving array utilization (associated with computation speed) and data reuse (associated with the number of memory access) arises.

Additionally, when the load mapping size significantly exceeds the array, tiling in both row and column directions towards the load mapping often introduces idle array part in the edge column or row of every tiling loop. To mitigate this, tasks from the next column or row can be brought in prematurely to fill the idle array to get a spatial cover. Therefore, the optimal one can be selected from two directions. As illustrated in Figure \ref{classify}, we can categorize these situations into four distinct cases:

\begin{table*}[htbp]
    \footnotesize
 \renewcommand{\arraystretch}{1.5}   
    \caption{Evaluated platforms' information.}
    \begin{center}
    \begin{tabular}{ |c |c | c | c | c| }     
         \hline
        \textbf{ } & \textbf{GTA} & \textbf{VPU-Ara} & \textbf{GPGPU-NVIDIA H100} & \textbf{CGRA-hycube} \\
         \hline
        Technology & 14nm & 14nm & 4nm & 28nm \\ 
         \hline
        Frequency & 1GHz & 250MHz & 1755MHz & 704MHz \\ 
         \hline
        Area & $0.35mm^2$ & $0.33mm^2$ & $814.00mm^2$ & $7.82mm^2$ \\ 
         \hline
        Cores & 4 & 4 & 528-tensor core  & 4x4  \\ 
         \hline
        Precision & \makecell{ INT8, INT16, INT32,\\ INT64, BP16, FP16,\\ FP32, FP64 } & \makecell{ INT8, INT16, INT32,\\ INT64, BP16, FP16,\\ FP32, FP64 } & \makecell{FP64, TF32, FP32, INT32, BP16\\, FP16, FP8, INT8} & \makecell{ INT8, INT16, INT32,\\ INT64, BP16, FP16,\\ FP32, FP64 }\\ 
         \hline
    \end{tabular}
    \end{center}
    \label{baseline}
\end{table*}

\begin{itemize}
\item \textbf{Uncover 1}: The workload falls short of covering the array in two directions.
\item \textbf{Uncover 2, 3}: The workload exceeds the array size in either the row or column direction, but the total size does not fully cover the entire array.
\item \textbf{Cover 2, 3}: The workload exceeds the array size in either the row or column direction, achieving full coverage of the entire array.
\item \textbf{Cover 1}: The workload exceeds the array size in both directions. The tiling placement could be in direction of Lateral or Vertical.
\end{itemize}

 We define this approach as dataflow pattern matching, resulting in a dataflow scheduling space. Because of multitude of scheduling schemes, we could prioritize choices based on a comprehensive priority strategy, wherein diverse outcomes are normalized, and the preference is given to the one with the least sum of squares.

\section{METHODOLOGY}
\subsection{Implementation}

Our architecture is based on the open source vector processor Ara~\cite{Ara2020}. The RTL design is synthesized using Synopsys Design Compiler and the SAED 14nm library to evaluate the chip area and energy of our architecture. At the same time, we also synthesize Ara under the same conditions for comparison. The maximum clock frequency of reference design is only approximately 250 MHz under our technology library. After replacing the original computing units with MPRA, the design can be synthesized at 1 GHz. Obviously, it is very advantageous in terms of timing under the design of basic PE in 8-bit. The lane with $8\times8$ MPRA can be implemented using only 60.76\% of the original lane area and cover all precision. Adding additional processing units for floating-point numbers, the overall area is about the same as that of the original lane. Due to the reuse of existing structures, the control and other logic have only 6.06\% area overhead over original Ara's setting 4 lanes.

\begin{figure}[htbp]
\centerline{\includegraphics[width=2.5in]{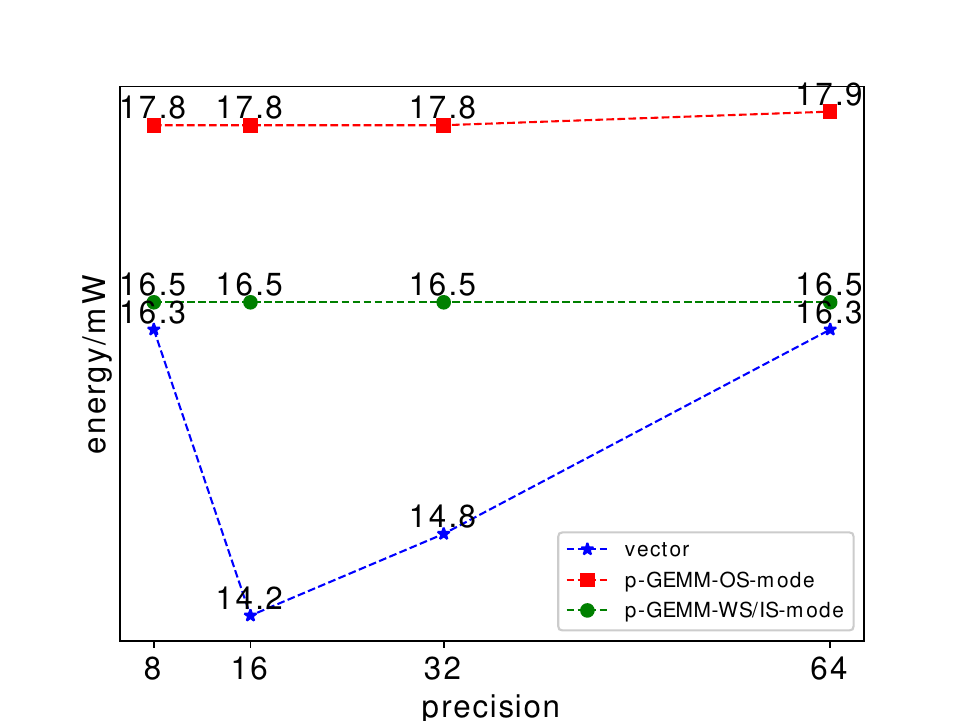}}
\caption{MPRA's energy when executing different mode.}
\label{Power breakdown}
\end{figure}

We measure the power consumption of MPRA under different precision and operation modes as shown in Figure \ref{Power breakdown}. The abscissa refers to the corresponding integer and floating point precision. The clock constraint is set at 250 MHz. It can be observed that their energy consumption is approximately the same. Although MPRA's average energy consumption is a little higher than original lane's computation unit, it can significantly reduce the energy efficiency of memory access.\par

\subsection{Workload} 

 \begin{table}[htbp]
    \footnotesize
    \renewcommand{\arraystretch}{1.2}         
    \caption{Size and precision of p-GEMM operators in various workloads.}
    \begin{center}
    \begin{tabular}{ |c | c | c| }     
         \hline
        \textbf{Workload} & \textbf{Description} &\textbf{Precision} \\
         \hline
        BNM & \makecell{Big Numbers Multiplication \\in Scientific Computing and Encryption Field} &INT64\\ 
         \hline
        RGB & SRGB2XYZ in Image Processing &INT8\\ 
         \hline
        FFE & FFE in Audio Processing &INT16\\ 
         \hline
        MD & \makecell{Matrix Decomposition\\in Mathematical Analysis}&INT32\\ 
         \hline
        PCA & PCA in Data Analysis&FP64\\ 
         \hline
        ALT & Alexnet Training in ML&FP32 \\
         \hline
        FFL & GPT3-Feed-Forward Layers in ML &BP16 \\
         \hline
        ALI & Alexnet Inference in ML&INT8 \\
         \hline
        Nerf & Nerf in ML&FP32 \\
         \hline
    \end{tabular}
    \end{center}
    \label{table_workload}
\end{table}

As listed in Table \ref{table_workload}, we select important tensor applications in various precision that are prevalent in various domains, and decompose them into p-GEMM and vector operators for execution.

\begin{figure*}[htbp]
\centerline{\includegraphics[width=7in]{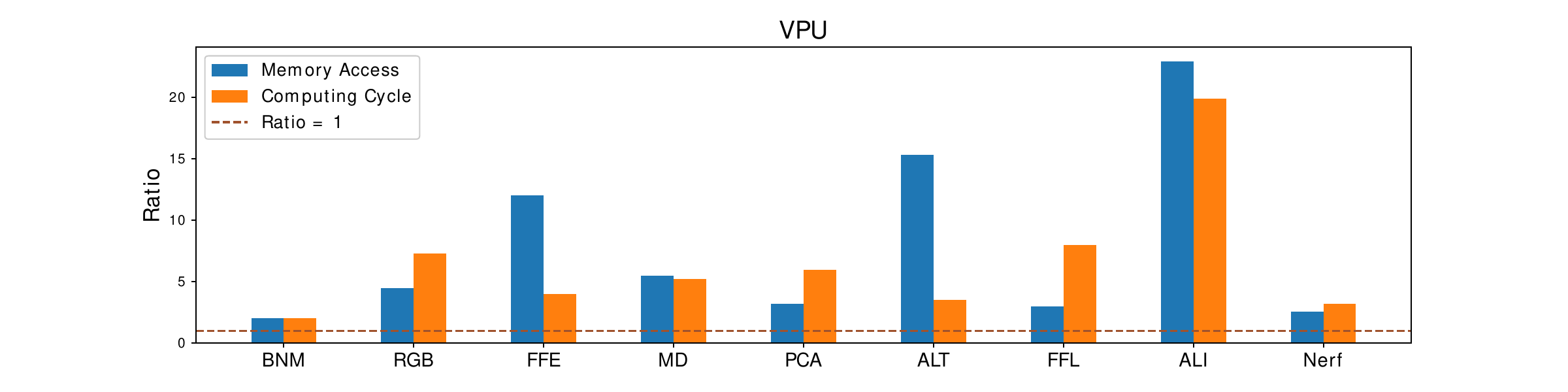}}
\caption{Comparisons with original VPU.}
\label{VPU_bar}
\end{figure*}

\begin{figure*}[htbp]
\centerline{\includegraphics[width=7in]{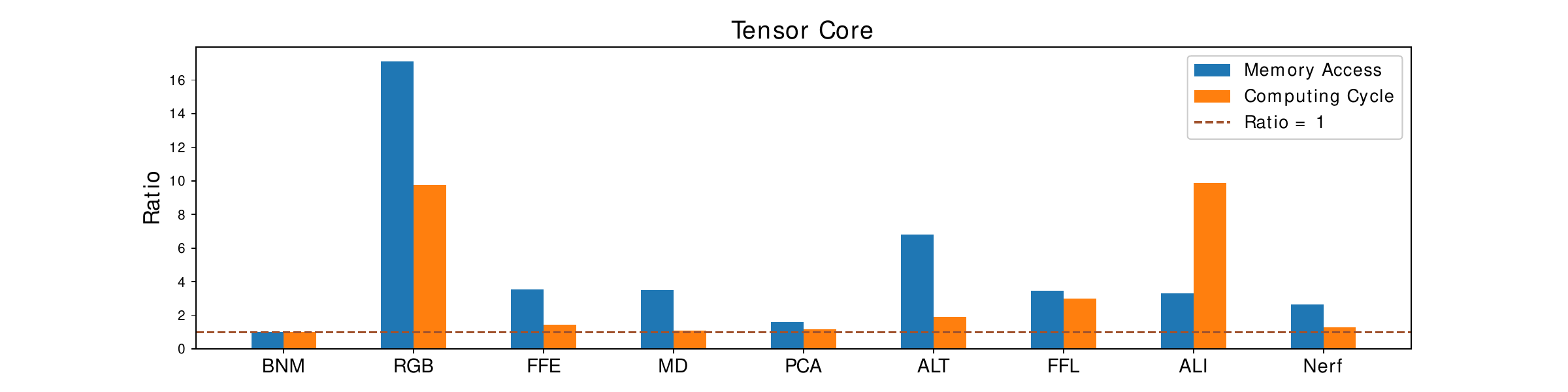}}
\caption{Comparisons with original GPGPU.}
\label{Tensor Core_bar}
\end{figure*}

\subsection{Baseline} 

GTA can be configured to operate in various scenarios, which vary in terms of memory configurations and array sizes. Therefore, our focus is specifically on two most important aspects, computing cycle and memory access, for core computing architecture. We assume the same clock frequency and configure different number of MPRA to match the same area according to technology library. The baseline is lited in Table \ref{baseline}:

\begin{itemize}
\item \textbf{VPU}: The vector units are parallel precision units essentially. Therefore we take Ara as an example and set the lane with all kinds of precision units. 

\item \textbf{GPGPU}: We choose the NVIDIA's newest Hopper ~\cite{choquette2023nvidia}. It involves Tensor Core and CUDA Core. For precision that Tensor Core cannot support, we use the closely higher precision.

\item \textbf{CGRA}: We choose CGRA with all kinds of precision to compare performance with GTA. We choose the hycube architecture~\cite{karunaratne2017hycube}.
\end{itemize}

We develop cycle-accurate simulators, based on scale-sim~\cite{samajdar2020systematic}, CGRA simulator morpher~\cite{morpher-woset2022}, VPU simulator~\cite{ramirez2020risc} and GPU simulator~\cite{khairy2020accel,luo2024benchmarking}, of our architecture and other baselines to get the performance statistics. We verify the GTA's simulator against our verilog implementation. 


\section{EVALUATION}

\subsection{Schedule Analysis}

\begin{figure}[htbp]
\centerline{\includegraphics[width=2.5in]{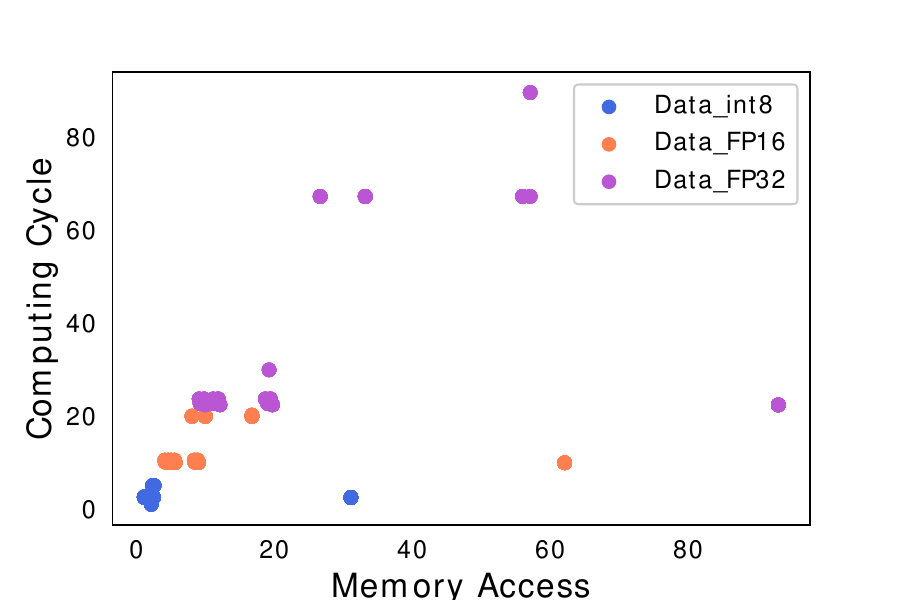}}
\caption{The scatter diagram of the scheduling cases with computing cycles and memory access index.}
\label{scatter}
\end{figure}

For the first time, we explore the mixed scheduling of precision and dataflow as shown in Figure \ref{scatter}. The values on both axes are the ratio to the minimum value gotten from all the available configurations. We choose one conv layer in Alexnet and set three kinds of precision used in case of real-world case. Due to the flexibility of our architecture and the exploration of scheduling space, we can display the statistical distribution of the scheduling cases. It is noteworthy that, unlike placing the precision units independently in other accelerators, different precision results in nonlinear distributions for the same operator.

\subsection{Compared with VPU}

The vector operators commonly encountered in every application. As shown in Table \ref{precision}, an increase in throughput is observed as precision decreases when compared to the original VPU. The bitwidth of mantissa in floating-point numbers is not multiples of 8 bits, so the gain is not an integer. Leveraging the reconfigurability of 8 bit-width computing units, the utilization is higher than the original design.\par

\begin{table}[htbp]
    \footnotesize
    \centering
    \renewcommand{\arraystretch}{1.3}         
    \caption{SIMD gains for all data types.}
    \begin{tabular}{cc|cc}
    \hline
    \textbf{Data Type} & \textbf{Throughput} & \textbf{Data Type} & \textbf{Throughput} \\ \hline
    INT8               & $8\times$     & BP16               & $16\times$    \\
    INT16              & $4\times$     & FP16               & $4\times$     \\
    INT32              & $2\times$     & FP32               & $3.56\times$  \\
    INT64              & $1\times$     & FP64               & $1.3\times$   \\ \hline
    \end{tabular}
    \label{precision}
\end{table}

\begin{figure*}[htbp]
\centerline{\includegraphics[width=7in]{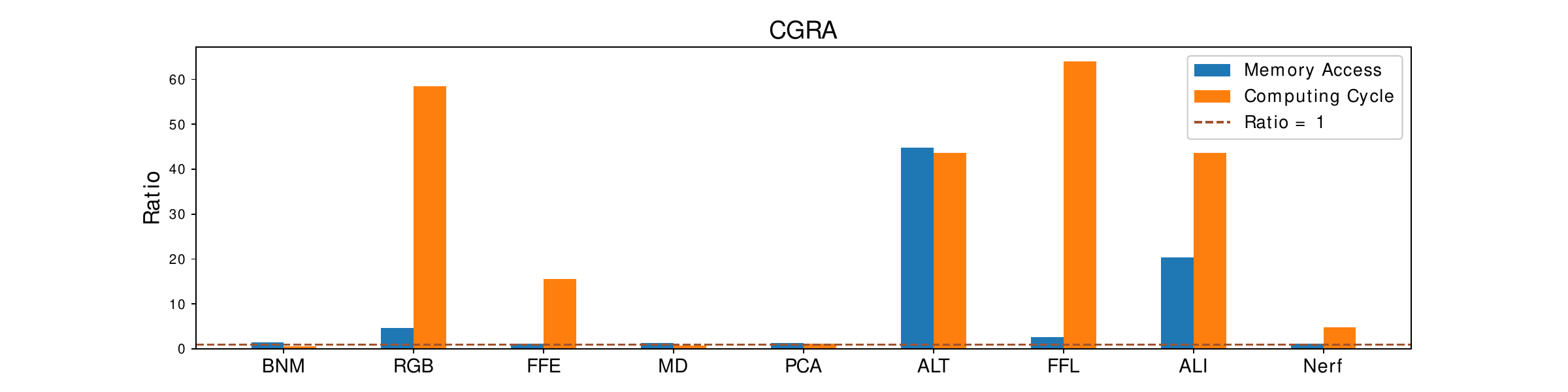}}
\caption{Comparisons with original CGRA in p-GEMM operators.}
\label{cgra}
\end{figure*}

Compared to systolic array architecture, the chaining technique in VPU exhibits weaker data reuse capability. Additionally, the number of computation units decreases with higher precision. Moreover, factors at the microarchitectural level, such as maximum vector length, also impose limitations on computational speed. As shown in Figure \ref{VPU_bar}, due to the area advantages brought by the additional computing units and the memory access advantages brought by systolic architecture, we can notice the savings in computational time and memory access.\par

The average memory access saving is $7.76\times$ and the average computational speedup is $6.45\times$. The results demonstrates strong overall performance. The comparison with VPU demonstrates the effectiveness of our structure in achieving notable generality in handling tensor operators in various domains.\par

\subsection{Compared with GPGPU}
The GPGPU consists of Tensor Core and CUDA Core. Tensor Core is only for accelerating GEMM .This work can be also regarded as the merger of Tensor Core and CUDA Core. Therefore, GTA have better area efficiency besides of the factor of precision. Furthermore, aiming to maximize the throughput, Tensor Core is consisted of small cube computing matrix multiplication, which requires large numbers of memory operations and high on-chip memory bandwidth. To get a fair comparison, we give the decomposed vector operator to cuda core and the p-gemm operator to tensor core.

As shown in Figure \ref{Tensor Core_bar}, there are still superiority of computing speed and the memory access. Due to the high throughput in high precision of Tensor Core, some performance remain modest, but there is a significant improvement in memory access. The average memory access saving is $5.35\times$ and the average computational speedup is $3.39\times$.\par

\subsection{Compared with CGRA}
CGRA realizes the flexibility for tensor operators, which use wordlevel reconfigurability and contain larger logic blocks and datapath-oriented interconnections. Therefore, CGRA is consisted of small arrays in physical implementation. As a result, they are relatively weak in acceleration and data reuse. It can be seen from the experimental data that high-precision computing units such as FP64 have a larger number of settings and can be on par with GTA in performance. But there are many PE in the idle state in the mapping. As shown in Figure \ref{cgra}, overall it still performs well. GTA is able to achieves $8.76\times$ memory efficiency and $25.83\times$ speedup over of CGRA.

\section{CONCLUSION}
Based on the similarity between matrix multiplication and precision multiplication and classification of tensor operators using p-GEMM and vector, We propose GTA, a general tensor accelerator combining of systolic array and VPU. GTA covers tensor operators of arbitrary precision. Besides, we explore the schedule space consisting of dataflow, precision and array resize. GTA achieves significant speedup and memory benefits compared
to other general accelerators.


\end{document}